\documentclass[aps,prb,superscriptaddress,twocolumn,floatfix]{revtex4-1}

\usepackage[utf8]{inputenc}
\usepackage[T1]{fontenc}
\usepackage{amsmath}
\usepackage{amssymb}
\usepackage{mathtools}
\usepackage{graphicx}
\usepackage[usenames,dvipsnames]{color}
\usepackage{bm}
\usepackage{hyperref}
\usepackage{url}
\usepackage{subfigure}
\usepackage{slashed,bbm}
\usepackage{dsfont}
\usepackage{setspace}
\usepackage{soul}
\usepackage{braket}
\usepackage{sidecap}

\sidecaptionvpos{figure}{c}

\definecolor{debianred}{rgb}{0.84, 0.04, 0.33}
\definecolor{checkcolor}{rgb}{0.44, 0.04, 0.83}

\renewcommand{\vec}[1]{\bm{#1}}

\newcommand\figinc[1]{\includegraphics[scale=0.76]{fig/#1}}

\begin{document}

\allowdisplaybreaks

%-----------------------------------
\title{Inherited and flatband-induced ordering in twisted graphene bilayers}
%-----------------------------------

\author{Lennart Klebl}
\author{Carsten Honerkamp}
\affiliation{Institut für Theoretische Festkörperphysik, RWTH Aachen University,
and JARA Fundamentals of Future Information Technology, Germany}

%-----------------------------------------
\date{\today}
%-----------------------------------------

%-----------------------------------------
\begin{abstract}
  The nature of the insulating and superconducting states in twisted bilayer
  graphene systems is intensely debated.  While many works seek for explanations
  in the few flat bands near the Fermi level, theory and a number of experiments
  suggest that nontwisted bilayer graphene systems do exhibit -- or are at least
  close to -- an ordered, insulating ground state related to antiferromagnetic
  ordering. Here we investigate in which ways this magnetic ordering scenario is
  affected by the slight twisting between the layers. We find that at charge
  neutrality the ordering tendencies of twisted systems interpolate between
  those of untwisted AA and AB stacked bilayers at intermediate temperatures,
  while at lower temperatures of the order of typical flat-band dispersion
  energies, the ordering tendencies are even enhanced for the twisted systems.
  The preferred order at charge neutrality still exhibits an antiferromagnetic
  spin arrangement, with ordered moments alternating on the carbon-carbon bonds,
  with an enveloping variation on the moiré scale. This ordering can be
  understood as inherited from the untwisted systems. However, even in the RPA
  analysis, the possible low-energy behaviors are quite versatile, and slight
  doping of one or more electrons per moiré cell can take the system into a,
  potentially flat-band induced, ferromagnetic phase.
\end{abstract}
%-----------------------------------------

%-----------------------------------------
\maketitle
%-----------------------------------------
The discovery of superconductivity in the vicinity of partially insulating
correlated-metal states in twisted bilayer graphene samples has triggered
numerous
experimental~\cite{cao2018correlated,cao2018unconventional,yankowitz2018tuning,%
Efetov}
and theoretical efforts. According to numerous theoretical
works\cite{doi:10.1021/nl902948m,Bistritzer12233,PhysRevB.86.155449,%
li2010observation,PhysRevLett.109.196802,doi:10.1021/acs.nanolett.6b01906,%
PhysRevX.8.031087,PhysRevX.8.031088,PhysRevX.8.031089},
in parts already before the discovery of superconductivity in these systems, the
slight twisting leads to a remarkable modification of the low-energy spectrum,
with 4 rather flat bands which are separated from the other bands quite well in
the case of magic twist angles. Many explanation
attempts\cite{PhysRevB.98.075154,roy2018unconventional,PhysRevB.97.235453,%
tang2018spin,PhysRevLett.121.217001,PhysRevX.8.041041,PhysRevB.98.245103,%
da2019magic,PhysRevB.98.241407,lin2019chiral,ClassenMoire}
involve enhanced electron correlation effects due to this condensed low-energy
spectrum. 

Ab-initio estimates for the interactions in and between the Wannier states of
the flat bands come up with quite large values exceeding the bandwidth of the
flat bands\cite{PhysRevX.8.031087,PhysRevX.8.031088}. However it can be expected
these large values will be screened down by the remnant $\pi$-band
spectrum\cite{Pizarro}. In general, it is believed that focusing on the flat
bands is legitimate as they are separated from the rest of the $\pi$-band
spectrum by a small energy gap of a few $\mathrm{meV}$s, but whether this is
sufficient to exclude decisive influence by the large rest of the spectrum, is
hard to know without well-defined calculations. 

In this paper, we avoid this uncertainty by working with the full $\pi$-bands.
On this larger energy scale, the interactions are smaller or comparable to the
band width. In fact, the strength of onsite and nonlocal interactions for
$\pi$-electrons has been a resurfacing topic over quite some time in solid state
theory\cite{BaeriswylMaki,Bursill,Wehling2011,Schueler}. More recently, the
interaction as a function of distance has been analyzed starting from ab-initio,
also taking into account environmental screening\cite{Roesner}. So, as a
starting point, the $\pi$-band model is quite well-defined and should allow one
to obtain almost quantitative insights, with the main uncertainties arising from
the theoretical treatment of the nonlocal moderate interactions within this
model.   A second reason for considering the full $\pi$-bandwidth is that a few
years ago, nontwisted, freely suspended bi- and trilayer graphene systems were
investigated intensely, as experiments showed the opening of a likely
interaction-induced gap of a few meV width at low
temperatures\cite{Velasco,Bao,Freitag,Veligura,Bao}. The theoretical
understanding is that this gap was likely due to the onset of antiferromagnetic
order in the layers\cite{Nilsson,Lang2012,SchererBi,SchererTri} with staggered
moments within one layer and unequal absolute magnitude on the two sublattices.
Note that in effective models for AB-stacked bilayer where only the two bands
touching the Fermi level are taken into account, this state comes in disguise as
intralayer ferromagnetic inter-layer antiferromagnetic
\lq{}layer antiferromagnet\rq{}
(LAF)\cite{ZhangJung,Throckmorton,Kharitonov},  as only one sublattice per
layer is considered. More recently, experiments on Bernal-stacked
multilayers\cite{Grushina,Morpurgo,Myhro} exhibited quite the same gap phenomena
with clear increase of the gapping temperature up to $100\,\mathrm{K}$ and,
not unexpectedly based on the band structure, an even-odd effect in the low-$T$
conductance in the number of layers\cite{Morpurgo}.  It should however be
mentioned that not all experiments on bilayers\cite{Yacoby,Weitz,Mayorov}
showed a full gap at low $T$ and also theoretically, there are other
options\cite{Throckmorton,Lemonik,ClassenPhon}. So it may well be some these
layered graphene systems are only on the verge of such ordering instability and
additional details of the samples or the environment decide about its actual
occurrence and type of order. 

Now, given the observation of insulating states in twisted bi- and multilayer
systems one may ask if these states are in any way related to the potential
ordering instability in the nontwisted systems. Hence, the idea of the current
manuscript is to use an approach that produces the AF instability as a likely
candidate for the nontwisted bilayer systems and to see what the changes are if
a slight twisting is introduced.  In fact, the occurrence of an
antiferromagnetic ordering on the C-C bond scale for the twisted bilayer is
already shown in self-consistent studies for onsite interactions in
Ref.~\onlinecite{Lado}. Here we ask if this instability is stronger in the
twisted system compared to the AB or AA bilayer and analyze the scenario in more
depth, e.g. regarding the consequences of doping.

In order to make this comparison, we set up the nontwisted and twisted unit
cells of equal size and derive the non-interacting band structure of the moiré
superlattice. Then we compute the relevant particle-hole diagrams at zero
momentum transfer, from which we can infer instabilities toward magnetic order
using the random phase approximation (RPA), which in this case can be understood
as a generalized Stoner theory. We compare the strength of the  magnetic
ordering as a function of the temperature and band filling and relate it to
nontwisted bilayer systems. Then we also study the ordering pattern within the
moiré unit cell and use this as estimate for the ordered moment below the
ordering transition. Inserting this as meanfield into the electronic dispersion
yields renormalized, split-up bands, which can be related to the experimental
observation of insulating states.

\section{Model}
\begin{figure} \centering
  \figinc{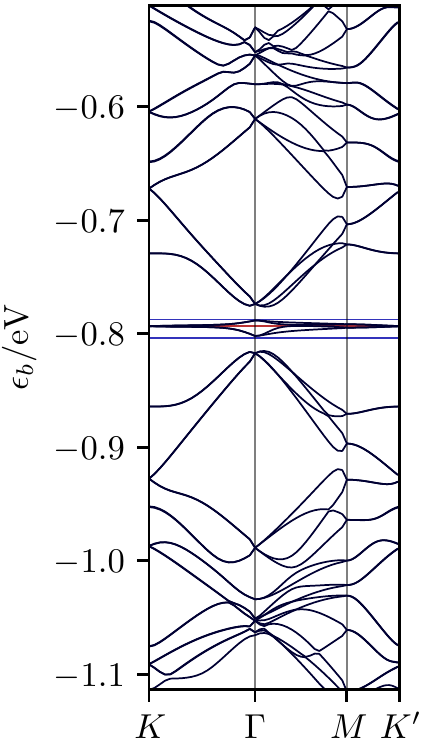}
  \figinc{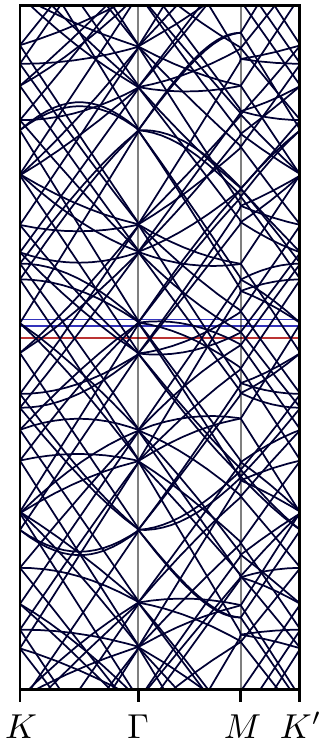}
  \figinc{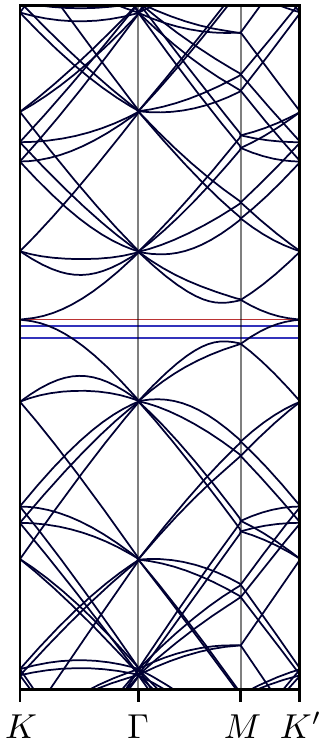}
  \caption{Band structure in the moiré Brillouin zone of the 31/32 super cell.
  In the left plot we show the bands for the twisted bilayer with
  $\theta = 1.05^\circ$, in the middle for the AA-stacked and in the right plot
  for the AB-stacked 31/32-system. The red lines are the three Fermi
  energies at charge neutrality, the blue lines are the other systems' Fermi
  energies as a reference. The bandwidth of the four flat bands for the twisted
  bilayer is approximately $16\,\mathrm{meV}$.}
  \label{fig:specplot}
\end{figure}

We start with an AA bilayer of carbon sites spanned by the two in-plane Bravais
lattice vectors $\vec{l}_1 = (\sqrt 3 /2, 3/2,0)$ and $\vec{l}_2 = ( \sqrt
3,0,0)$ and two sites per unit cell separated by the vector $(0,1,0)$, all
measured in units of the C-C bond length $a_0=0.142$ nm. The
\lq{}vertical\rq{} distance
between the two layers is given by\cite{Moon} the shift vector $0,0,d)$ with
$d=0.335$ nm$=2.36 a_0$.  The large unit cell of the twisted bilayer system is
obtained by defining two superlattice vectors $\vec{L}_1= n
\vec{l}_1+m\vec{l}_2$ and $\vec{L}_2$ rotated by 60 degrees with respect to
$\vec{L}_1$.  Then one of the two layers is rotated by the twist angle $\theta=\arccos
\frac{m^2+n^2+4mn}{2(m^2+n^2+mn)}$ around an AA lattice carbon site. For the
magic angle $\theta =1.05^\circ$ we use $n=31$ and $m=32$ (denoted
31/32-system). This produces a moiré unit cell of 11908 carbon sites. Next we
set up the Koster-Slater tight-banding Hamiltonian matrix as described in
Refs.~\onlinecite{doi:10.1021/nl902948m,Moon}, including the corrugation
detailed in the appendix of Ref.~\onlinecite{PhysRevX.8.031087}. 
From that we obtain for each $\vec{k}$ in
the folded Brillouin zone the band energies $\epsilon_b(\vec{k})$ and
eigenvectors $u_{ib}(\vec{k})$ for band $b$ and site index $i$ inside the moiré
unit cell. In Fig.~\ref{fig:specplot} we show the low-energy band structure in
the moiré Brillouin zone for the 31/32-system as well as the non-twisted AA and
AB bilayers using the same moiré supercell. One can clearly see the formation of
the flat bands in the twisted system in contrast with linear or quadratic wider
bands in the case of AA and AB bilayers (both with constant interlayer
distance). Due to the backfolding, the AA bilayer shows a numerous linear Fermi
level crossings in the small moiré Brillouin zone, while the AB bilayer exhibits
quadratic band crossing points at $K$ and $K'$. These are actually split up into
separate Dirac points on a low energy scale not visible in this plot, due to the
remote hoppings included in this calculation. 

Next, we consider Hubbard onsite interactions for electrons with opposite spins
residing on carbon sites $i$,
\begin{equation}
  H_U = U \sum_i n_{i,\uparrow} n_{i,\downarrow} \, .
\end{equation}
This interaction choice represents of course a major simplification with respect to the true long-range
Coulomb interactions that are present in the experimental system. Actually,
there are good cRPA estimates for the non-local Wannier-state interaction for
mono-and bilayer graphene, also taking into account the embedding in an
insulating substrate such as hBN\cite{Roesner}.  However, it is also known that
the primary instability tendencies are not affected qualitatively by the
non-local terms.  The latter terms only reduce the strength of the instability
and could actually be absorbed into a redefined effective onsite repulsion
$U^*$\cite{Schueler}. Here we take the point of view that we just compare the AF
instability trends for various bilayer configurations and by this avoid to pin
down the ordering strength quantitatively on an absolute scale.  Furthermore,
recent QMC\cite{Tang} and fRG\cite{SanchezdlP} studies have given evidence that
the leading instabilities are not changed by the non-local parts of the
interaction and that the dominant instability for sufficient low-energy density
of states is of AF-SDW type. This in turn can be well modeled with the
Hubbard-$U$-only interaction. From our comparison of non-twisted and twisted
systems we learn about how and how much the instability tendencies differ in
these systems.

\section{Effective RPA theory for the magnetic ordering}
The magnetic ordering in a large number of correlated electron systems can be
analyzed in simple terms by using the random-phase approximation (RPA) for the
magnetic susceptibility $\hat\chi(\vec{q},\nu)$. In this textbook-style
approach, the transverse (up-down) spin channel, the spin susceptibility in the
moiré unit cell can be written as
\begin{equation}
  \label{chiint}
  \hat\chi^{zz}(q) = \hat{\chi}_0 (q) \left[ \mathbbm{1} +
  U \hat{\chi}_0 (q) \right]^{-1} \, ,
\end{equation}
where the matrices with the hat-symbol have $N_M$ components, running over all
$N_M$ sites in the moiré unit cell. $\hat{\chi}_0(q)$ is the bare particle-hole
bubble, with the matrix elements
\begin{eqnarray}
  \label{chimatrix}
  \hat{\chi}_{0,ij} (\vec{q},\nu) &=& \frac{1}{N} \sum_{\vec{k},b,b'}
  \frac{n_F(\epsilon_b(\vec{k}+\vec{q}))- n_F(\epsilon_{b'}(\vec{k}))}{-i \nu +
  \epsilon_b(\vec{k}+\vec{q}) - \epsilon_{b'}(\vec{k}) } \nonumber \\ && \cdot
  \, u_{ib}(\vec k + \vec q) u^*_{jb}(\vec k + \vec q) u^*_{ib'}(\vec k )
  u_{jb'}(\vec k )
\end{eqnarray}
According to this formula, the calculation of the $N_M^2$ matrix elements
$\hat{\chi}_{0,ij}$ using the two internal summations $b,b'$  over $N_M$ bands
each would cause an effort $\mathcal O(N_M^4)$.  However, by computing
band-summed nonlocal Green's functions $G_{ij}(i \omega_n)$ and doing the
Matsubara sums numerically afterwards for products $G_{ij}(i \omega_n)G_{ji}(i
\omega_n)$, one can compute  $ \hat{\chi}_{0,ij} (\vec q,\nu=0)$ with effort
$\mathcal O(N_M^3 N_\omega)$ with $N_\omega \sim 500$ positive Matsubara
frequencies on an appropriately chosen grid. 

A divergence of the susceptibility (\ref{chiint}) is obtained when an eigenvalue
of $\hat{\chi}_0(\vec q, \nu)$ becomes $-1/U$, e.g. when the temperature $T$ is
lowered or the interaction strength $U$ is increased.  This criterion is the
generalization of the well-known Stoner criterion for magnetic ordering. Note
that, in order to keep the numerical effort limited,  we compute the bubble
diagram using 8 or 18 points in the irreducible moiré Brillouin zone. We have
checked that the contributions of the lowest Matsubara frequencies do not change
significantly if we use up to 72 points. Furthermore, we ignore possible
selfenergy effects on the internal lines of the RPA bubbles.

Alternatively, we can Hubbard-Stratonovitch-decouple the Hubbard interaction in
the functional integral formalism using a site-dependent  magnetization
$m_i(\tau)$. Then the fermions can be integrated out. The quadratic term in the
magnetization then reads (generalizing, e.g., Ref.~\onlinecite{Hertz}, and using
$q=(\vec{q},i\nu)$)
\begin{equation}
  \label{maction}
  S^{(2)}=  \frac{U}{4} \sum_{ij \atop q } m^*_i(q) \left[ \mathbbm{1} +
  U \chi_{ij} (q) \right] m_j(q)
\end{equation}
Again, at sufficiently low temperatures $T$ or for large enough interaction
strength $U$ and for $\nu=0$, the hermitian matrix in the square brackets will no
longer be positive definite and at least one eigenvalue $\lambda_0$ with
eigenvector $m^{(0)}_i$ will become smaller than zero. Then one can expect a
nonzero magnetization $m_i \propto m^{(0)}_i$ to develop spontaneously. This
onset of magnetic ordering is equivalent to the above-mentioned criterion for
the divergence of the interacting susceptibility.  It first happens at $\nu=0$,
as the particle-hole bubble is largest then. In the fermionic action, $m_i$
couples linearly to the spin-$z$ component of the electrons. This can cause a
gap in the spectrum around the Fermi level, depending on the dispersion and band
filling.  Previous studies\cite{Honerkamp2008,SchererBi} of mono- and nontwisted
bilayer graphene have shown that the preferred ordering is at $\vec{q}=0$, with
a staggered eigenvector with a sign change between the nearest neighbors on the
honeycomb lattice and between the layers in the case of the AB-stacked bilayer.
The relative size of the ordered moments in meanfield and
QMC-studies\cite{Lang2012} agree qualitatively with that found in the leading
eigenvector in $\hat \chi$ in the fRG\cite{SchererBi}, i.e. it is also larger on
those sites that are not connected by the interlayer hopping. This is to be
expected, as the weight of the bands touching at the Fermi level is higher on
these sites than on the sites coupled by the interlayer hopping. 

\section{Critical interaction strength for instability}
\begin{figure} \centering
  \figinc{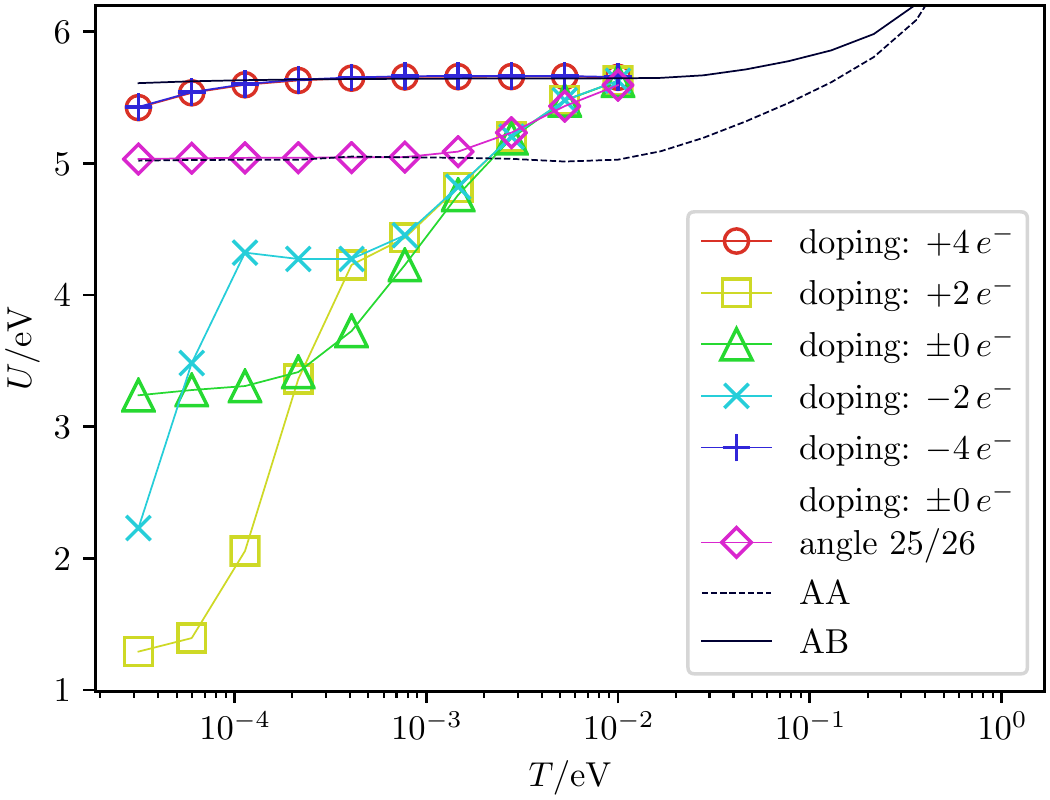}
  \caption{Critical RPA onsite interaction strength vs. temperature for the
  $\vec{q}=0$ instability in the AA (dashed) and AB (solid line) bilayer without
  twist, and for the 1.05-degree twisted system and different fillings of the
  flat bands around the Fermi level, including remote hoppings as given in
  Ref.~\onlinecite{PhysRevX.8.031087}. In addition, the 25/26 twisted system
  ($\theta=1.30^\circ$) at charge neutrality is shown.}
  \label{fig:ucvst}
\end{figure}

In a first step we reanalyze the instability tendencies of the non-twisted
bilayers with AB and AA stacking, by tracking for which parameters we find the
divergence of the interacting susceptibility \ref{chiint} described in the last
section. For the \lq{}plain-vanilla\rq{} version with nearest-neighbor hopping
only and without remote hoppings, the SDW instability was studied in detail e.g.
in Ref.~\onlinecite{Lang2012}, using RPA and QMC. Here we add to this by
including the remote hopping processes as described by the tight-binding
modeling of Ref.~\onlinecite{PhysRevX.8.031087}. These additional hoppings
remove quadratic band crossing points in the case of the AB stacking and also
alter the dispersion in the AA case. Hence, we no longer find instabilities at
infinitely small $U$ as in some previous studies. The critical onsite-$U$s are
shown for a range of temperatures in Fig.~\ref{fig:ucvst}.

As mentioned above, the possibility of a antiferromagnetic ordering on the C-C
bond scale for the twisted bilayer was already shown in selfconsistent studies
for onsite interactions in Ref.~\onlinecite{Lado}. Here we ask if this
instability is stronger in the twisted system compared to the AB or AA bilayer.
In the same Fig.~\ref{fig:ucvst} we hence add the $U_c$-values for the
$\theta=1.05^\circ$ twisted system. For the charge-neutral system we will show
below that the instability is also essentially of antiferromagnetic type. We
observe that for higher $T$ the $U_c$-values lie between those for AA and AB
stacking, while they deviate significantly to lower $U$ at smaller $T$s for the
charge-neutral twisted bilayer. This shows that the twisted system at charge
neutrality is even more susceptible to a magnetic instability than the
non-twisted systems. The deviation occurs on temperatures scales of
$1\,\mathrm{meV}$, or $10\,\mathrm{K}$, which is the temperature range when the
flat bands are resolved. Hence this finding is not unexpected. Consistently,
somewhat away from the magic angle, at $\theta=1.30^\circ$, the enhancement is
less marked.

Interestingly, as shown in Fig.~\ref{fig:ucvst}, the $U_c$ values for the $\pm
2$-doped systems are even lower than the ones for the charge-neutral system. We
see below that the ordering pattern is also different for those cases, but we
cannot offer a simple intuitive explanation for this here. In any case it
supports the statement that the flat bands of twisted bilayer graphene exhibit a
rather rich and versatile physics.

\section{Spatial structure of the magnetic ordering}
Next we discuss the type of magnetic order that is suggested by this RPA
analysis. This becomes clear from the spatial structure of the eigenvector that
belongs to the leading eigenvalue of the instability.

\begin{figure}
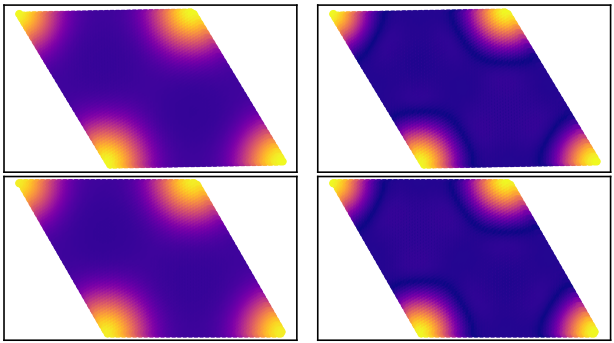
 \centering
  \figinc{{beta1.00000e+02_fill+0_abs}.pdf}
  \figinc{{beta2.44944e+03_fill+0_abs}.pdf}
  \caption{RPA data at charge neutrality: absolute value of the leading
  eigenvector proportional to the magnetic order parameter in the two-layer
  rhomb-shaped unit cell (blue to yellow: smallest to largest absolute value,
  upper plots: upper (rotated) layer, lower plots: lower layer).  The AA-regions
  are in the corners of the rhomb, the AB- and BA-region on the diagonal from
  the upper left to the lower right corner at one and two thirds distance.  Left
  plot: intermediate temperature $T=10\,\mathrm{meV}$, right plot: low
  temperature $T=0.408\,\mathrm{meV}$.}
  \label{fig:dope0abs}
\end{figure}

\begin{figure}
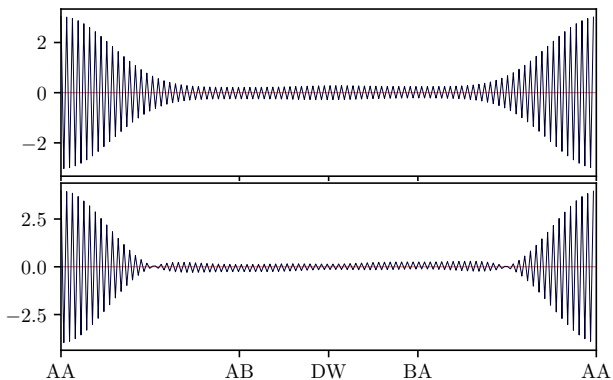
 \centering
  \figinc{{beta1.00000e+02_fill+0_sgn_linecut}.pdf} \\[-12pt]
  \figinc{{beta2.44944e+03_fill+0_sgn_linecut}.pdf}
  \caption{RPA data at charge neutrality: line cut of the leading eigenvector
  proportional to the magnetic order parameter through the rhomb-shaped unit
  cell in one layer, starting in the AA-regions in the corners of the rhomb
  through the AB- and BA-region on the diagonal to the next AA-region.  Upper
  plot: intermediate temperature $T=10\,\mathrm{meV}$, lower plot: low
  temperature $T=0.408\,\mathrm{meV}$ with an additional node of the order
  parameter near the AA regions.}
  \label{fig:dope0line}
\end{figure}

First, let us consider the charge-neutral system. In Fig.~\ref{fig:dope0abs} we
plot the absolute value of the eigenvector in the moiré unit cell for two
temperatures. We see that the eigenvector is larger in the AA regions in the
corner of the rhomb-shaped unit cell. At the higher $T$ the eigenvector is
somewhat more extended away from the AA regions compared to the data at lower
$T$, in particular along the diagonal which separates the AB region from the BA
region.  In the plots below, in Fig.~\ref{fig:dope0line}, we show lines cuts of
the eigenvector from the AA region through AB and BA regions to the other AA
region on the other side of the unit cell in one layer. We clearly observe the
sign oscillations, which identify the instability as toward antiferromagnetic
(AF) order with opposing signs of the order parameter on A and B sublattice in
the same layer. This order also occurs in homogeneous fashion for the nontwisted
AA and AB bilayers. It appears that the twisted system just {\em inherits} this
order parameter from the non-twisted bilayers, with additional modulation owing
to the inhomogeneous distribution of the density of states at lowest energies.

The more precise comparison between the two eigenvectors at different
temperatures reveals that at the lower $T \lesssim 2\,\mathrm{meV}$, which is of
the order of the van Hove $M$-point energy in the  flat-band dispersions, the
staggered magnetic order changes its registry along the path from the AA to the
AB or BA region.  This is clearly visible as a node in the absolute value of the
order parameter.  Besides the overall modulation through the unit cell, this is
another alteration of the ordering in the twisted systems, indicating that the
flat bands induce additional physical complexity.

\begin{figure}
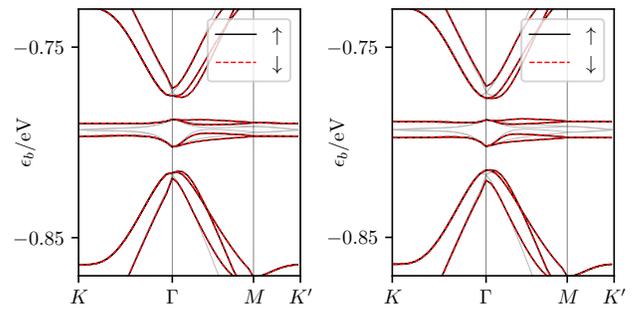
 \centering
  \figinc{{meanfield0.5_beta1.00000e+02_fill+0}.pdf}
  \figinc{{meanfield0.5_beta3.16228e+04_fill+0}.pdf}
  \caption{Magnetic band structure obtained by inserting the site- and
  spin-dependent leading eigenvector multiplied by $U/4$ (with $U \approx
  2\,\mathrm{eV}$) into the diagonal of the
  hopping Hamiltonian. Left plot: intermediate temperature $T=10\,\mathrm{meV}$,
  gap size $\Delta \approx 6.70\,\mathrm{meV}$, right plot: low temperature
  $T=0.03\,\mathrm{meV}$, gap size $\Delta \approx 8.36\,\mathrm{meV}$. The path
  goes through the reduced Brillouin zone of the moiré lattice. The nonmagnetic
  bands are indicated by the faint lines.}
  \label{fig:dope0bands}
\end{figure}

\begin{figure}
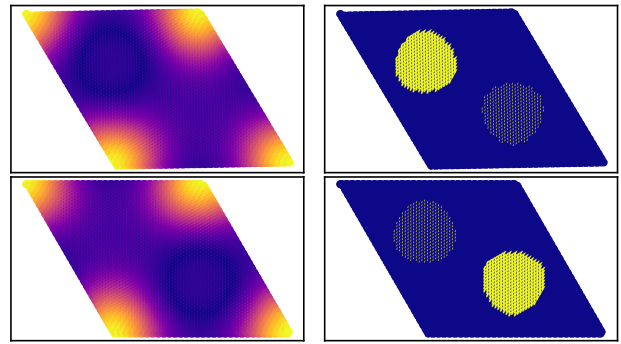
 \centering
  \figinc{{beta4.64159e+03_fill-2_abs}.pdf}
  \figinc{{beta4.64159e+03_fill-2_sgn}.pdf}
  \caption{Absolute value (left) and sign (right) of the leading eigenvector for
  the $+2$ doped system and temperature $T=0.215\,\mathrm{meV}$.}
  \label{fig:dope-2eigenvector}
\end{figure}

In order to understand the impact of the magnetic order on the electronic
spectrum, we insert the leading eigenvector of the susceptibility as our guess
for the meanfield that occurs in the diagonal of the hopping Hamiltonian of the
electrons. At charge neutrality, we easily find a gap opening at the Fermi level
due to the ordering.  Two cases at different temperatures, one for the nodeless
AF ordered state and one for the nodal state a lower $T$  are shown in
Fig.~\ref{fig:dope0bands}. We expect that a self-consistent solution of the
magnetic meanfield theory would give similar spectra. On the quantitative
side, our gaps come out an order of magnitude larger than the activation
energies observed experimentally\cite{Efetov}. Note however that we just
inserted $U/4$ times the eigenvector into the Hamiltonian as a
na\"i{}ve guess. No attempt to obtain a selfconsistent solution was made.
Furthermore, various fluctuations and renormalizations may influence the actual
gap size.  Qualitatively, our data supports the expectation that the AF ordered
states should be  gapped and hence exhibit insulator-like transport physics.  
Next we move to the doped system with $\pm n$ and $|n|\le 4$ electrons per moiré
unit cell.  Here the leading eigenvector at low $T$ and doping $+2$ is shown in
Fig.~\ref{fig:dope-2eigenvector}.

\begin{figure} \centering
  \figinc{{beta4.64159e+03_fill-2_sgn_linecut}.pdf}
  \caption{Line cut of the leading eigenvector for the $+2$ doped system. The
  path is again diagonal through the rhomb-shaped unit cell from the top left
  to the bottom right corner.}
  \label{fig:dope-2linecut}
\end{figure}

\begin{figure} \centering
  \figinc{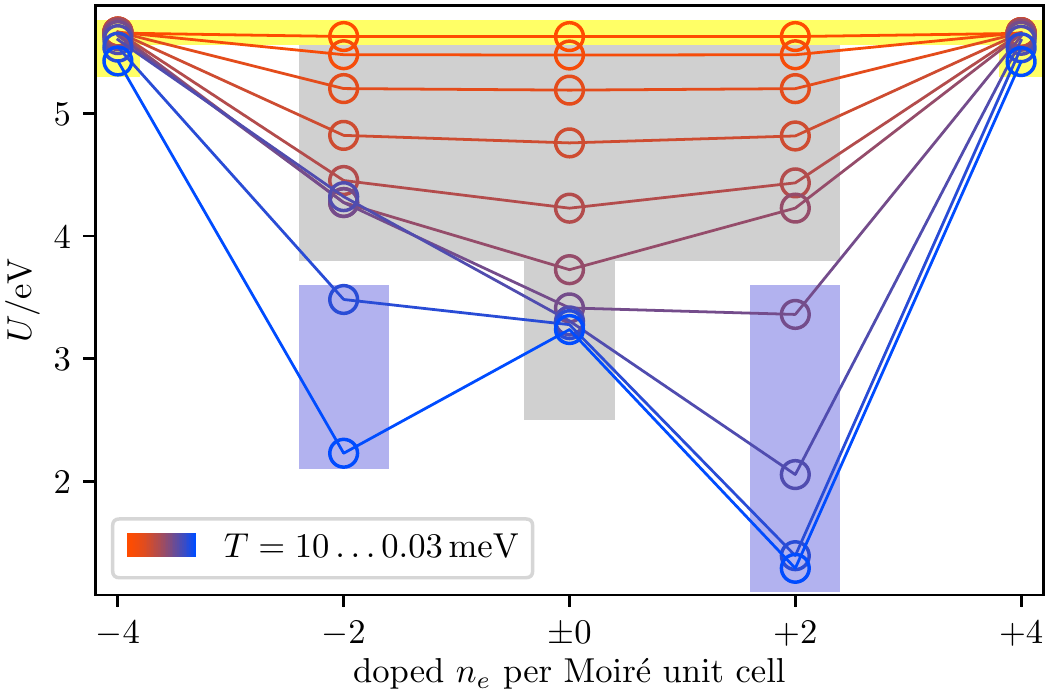}
  \caption{Tentative magnetic RPA phase diagram for the filling range of
  even integer doping of up to $\pm 4$  electrons per moiré unit cell. The lines
  show the threshold interaction strengths for magnetic ordering for different
  temperatures.  Lower values correspond to increased instability toward
  ordering. The color shading indicates the type of magnetic order -- yellow:
  nodeless antiferromagnetic order, grey: antiferromagnetism with real-space
  node in order parameter, blue: ferromagnetic order.}
  \label{fig:PD3132}
\end{figure}

We observe that while there is still some staggering on the eigenvector along
the trace through the unit cell, the overall sign of the eigenvector is the same
in most of the unit cell  except for islands with oscillating sign near the AB
and BA regions. This means that the leading instability is of ferromagnetic (FM)
type now, again with strongest weight in the AA regions. Again, we can insert
the eigenvector as order parameter into the hopping Hamiltonian. It turns out
that the spectra at $\pm 2$ become gapped rather readily but that it is hard to
induce a gap for odd-integer dopings via this simple route. Here we would have
to invoke a stronger coupling or additional mechanisms beyond our present
scheme.

\begin{figure}
  \figinc{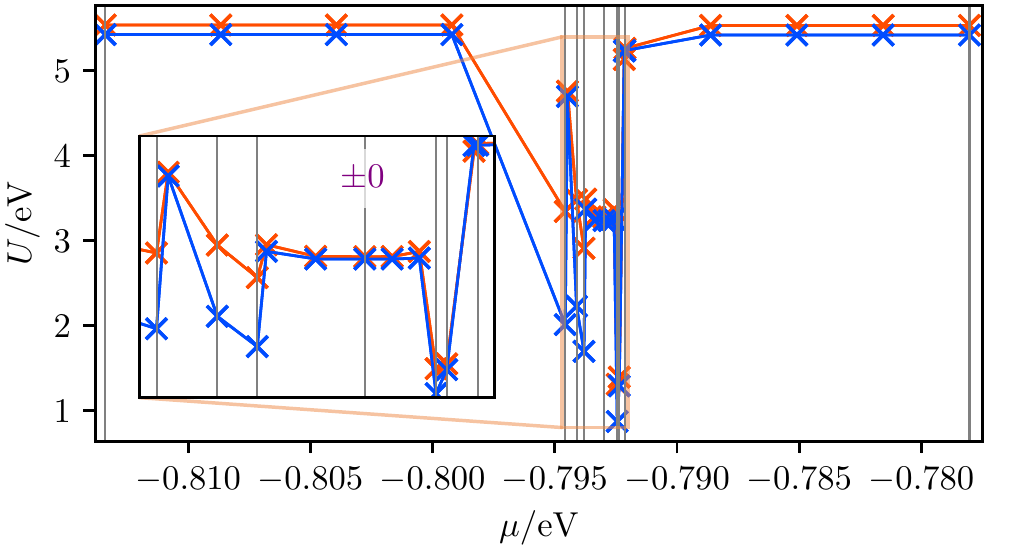}
  \caption{Critical onsite interaction strength $U$ versus chemical potential
  $\mu$, indicating the instability tendency for dopings in the filling range
  $\pm4$ electrons per moir\'e unit cell. The vertical lines indicate integer
  fillings $-4,-3, \dots , 4$ with charge neutrality (filling 0) at
  $\mu\approx0.793\,\mathrm{eV}$. Red line: temperature $T=0.06\,\mathrm{meV}$,
  blue line: temperature $T=0.03\,\mathrm{meV}$. The inset shows the same data
  on an enlarged $\mu$-scale.  All integer dopings $-3$ to $+2$ electrons per
  moiré unit cell show an increased magnetic ordering tendency.}
  \label{fig:uc-dense-mu}
\end{figure}

Notably, at somewhat higher temperatures, we still find the antiferromagnetic
solution, and only at lower $T$ the FM eigenvalue becomes strongest. Again this
supports the idea that the flat-band physics is quite rich. The higher-$T$ state
is the inherited order from the nontwisted system, which   is dominated by the
bulk of the spectrum, while at lower $T$ the flat bands add additional features
and alter the ordering pattern to ferromagnetic alignment.

\section{Tentative phase diagrams}
We can summarize our findings in a tentative RPA phase diagram for the magnetic
ordering ordering in the magic-angle bilayer system. This is shown for selected
even integer dopings in Fig.~\ref{fig:PD3132} for a larger number of
temperatures.  We also include dopings other than $\pm 2$ in Fig.
\ref{fig:uc-dense-mu}, where we scan the critical interaction strength as a
function of the chemical potential $\mu$ for two low temperatures.  According to
this data, there is an enhanced ordering tendency for the filling range between
$-3$ to $+2$ electrons around charge neutrality at filling 0. Now, if the
relevant onsite interaction strength $U$ would be around 4 or 5eV, as argued in
Ref.~\onlinecite{Roesner}, this would result in a window for magnetic order
between fillings $-3$ and $+2$ electrons per moiré unit cell. Then, the fillings
0 and $\pm 2$ should be insulating, as then the Fermi level would lie between
the respective bands. For other fillings, including $\pm 1$ and $+3$ electrons
per moiré unit cell, one should have a metallic state, potentially with a
magnetically ordered background. Of course, additional effects such as increased
coupling to the order parameter or  magnetic disorder could render these states
insulating as well. Outside the window $-3 $ to $+2$ electrons, we do not find
amplified ordering tendencies. In particular, within this simple picture, we
cannot explain the potential insulating state  at $+3$ electrons, because the
interaction tendency is not strongly enhanced there. The insulator at $\pm 4$ is
believed to be a trivial band insulator, which remains unchanged by our
analysis.

\section{Discussion and conclusions} In this random-phase approximation (RPA)
study of the full $\pi$-band spectrum, we have compared the instability tendency
of the twisted graphene bilayer to the untwisted AB- and AA-stacked version. The
main goal was to determine the ground state order, and from that we derived
estimates of the low-energy properties, with the goal to understand the nature
of the insulating states in magic-angle twisted bilayer graphene.

We found that at intermediate temperatures above~\mbox{$\sim 20\,\mathrm{K}$}
the instability strength of magic-angle twisted bilayers at charge neutrality
toward antiferromagnetic (AF) ordering lies close to that of the non-twisted
AB-stacked or AA-stacked systems.  For lower temperatures, the flat bands of the
twisted systems strongly enhance the RPA instability tendency, and the
corresponding threshold interaction strengths become quite small and fall below
the theoretical estimates for the effective onsite interaction in standard
graphene systems\cite{Wehling2011,Roesner,Schueler}.  Thus it is very likely
that the twisted systems may indeed order in this way at low $T$, at least
locally.  We can understand this AF instability {\em inherited} from the
nontwisted AB bilayers.  Note that various experiments showing low-temperature
gap openings in freely suspended Bernal(AB)-stacked graphene were interpreted in
terms of AF ordering\cite{Velasco,Morpurgo}.  Of course, the embedding in hBN of
the twisted system may cause some additional screening, but in general the
change in the short-range interaction parameters due to the hBN will be
limited\cite{Roesner}. Hence it is not implausible to consider the possibility
that the twisted bilayers order in an analogous way even if they are
encapsulated in hBN. It would be very interesting to search for the staggered
and unit-cell modulated spin correlations, potentially with spin-resolved
scanning tunneling techniques.  Away from charge neutrality, the main magnetic
ordering tendency turned out to be ferromagnetic (FM) when the temperature was
lower than the typical van-Hove-singularity flat-band dispersion energies. This
points to the impact of the flat bands in inducing FM correlations instead of
the AF order. Of course, it would be interesting to understand the relation of
this effect to known cases of flat-band ferromagnetism\cite{Mielke1993,Hlubina}.

Discussing potential caveats of our approach, one should mention that the RPA
used here clearly runs the danger of overestimating the ordering tendencies.
Channel coupling and selfenergy corrections, as well as non-local interactions
will reduce the energy scales for ordering and increase the threshold
interaction strength needed for the instabilities. We are currently working to
implement a functional renormalization group study of these systems, in order to
asses these issues more thoroughly. First results support the RPA picture.
Moreover, the differences between RPA and more controlled approaches in this
context are known to be limited from studies of nontwisted
systems\cite{Lang2012}.  On the positive side, compared to many studies focusing
on just four low-energy bands, our study is far more quantitative as it does not
ignore a large portion of the degrees of freedom. If the flat low-energy bands
show a different physics from that of the nontwisted layers, it should be
visible.  Indeed, we found that small doping with just one or two electrons per
unit cell leads to a qualitatively different, ferromagnetically ordered ground
state.

We furthermore argued that a (partial) sequence of insulating states can be
understood by inserting the corresponding suggested orders as meanfields, at
least for the charge-neutral and the $\pm 2$-doped cases. There, if the order
parameter is large enough, the flat bands split up such as to give gaps for
integer band fillings. If the density is away from these even integer fillings,
the split-up bands of the magnetically ordered state are partially filled, and
the system remains metallic. This opens windows for superconductivity in between
the integer fillings. Note that our current approach is not suitable to search
for superconductivity, but at least it allows one to understand how metallic and
insulating situations can arise as a function of he electron density in the flat
bands. The picture given here would mean that the superconductivity arises
away from integer fillings in addition to or in the background of the magnetic
ordering suggested by the RPA. The mechanism would then either be phononic or
electronic, by the fluctuations of the magnetic order. As the magnetic order
appears to be doping-dependent, there is the actual possibility of distinct
superconducting states for different dopings. As the magnetic order breaks
time-reversal and spin rotational symmetry, the superconducting states could be
unconventional, even if they are mediated by phonons. Exploring these
possibilities will be the content of future work.

We also commented on the fact that our current study cannot readily explain
insulating states a odd integer band fillings $\pm 1$ and $+3$. For these, one
may have to include additional symmetry breakings of self-energy effects. In
addition to that, charge redistributions\cite{GuineaWalet} and order at nonzero
wavevectors varying on the moiré scale\cite{Lado} can lead to additional physics
not accounted for here. It will be interesting to study how these effects add to
the  \AA ngstr\"om-scale magnetic order found here.

We thank L. Classen, A. Honecker, J. Lado, A. MacDonald, M. Morgenstern, and M.
Scherer for discussions.  The German Science Foundation (DFG) is acknowledged
for support through RTG 1995, and JARA-HPC for granting computing time.

\bibliography{Bib_moire}

%------------------------------------------------
\end{document}